\documentclass[sigconf]{acmart}

\usepackage{booktabs} 
\usepackage[normalem]{ulem}
\usepackage{multirow}
\usepackage{multicol}
\usepackage[para]{footmisc}
\useunder{\uline}{\ul}{}
\usepackage [english]{babel}
\usepackage [autostyle, english = american]{csquotes}
\MakeOuterQuote{"}

\hypersetup{
  bookmarks=true%
,bookmarksnumbered=true%
,hypertexnames=false%
,breaklinks=true%
,colorlinks=true%
,linkcolor=black%
,citecolor=black%
,urlcolor=black%
}

\copyrightyear{2018}
\acmYear{2018}
\setcopyright{acmcopyright}
\acmConference[SIGIR '18]{The 41st International ACM SIGIR Conference on Research and Development in Information Retrieval}{July 8--12, 2018}{Ann Arbor, MI, USA}
\acmBooktitle{SIGIR '18: The 41st International ACM SIGIR Conference on Research and Development in Information Retrieval, July 8--12, 2018, Ann Arbor, MI, USA}
\acmPrice{15.00}
\acmDOI{10.1145/3209978.3210124}
\acmISBN{978-1-4503-5657-2/18/07}
\definecolor{tealblue}{rgb}{0.21, 0.46, 0.53}
\definecolor{wildstrawberry}{rgb}{1.0, 0.26, 0.64}
\definecolor{ao(english)}{rgb}{0.0, 0.5, 0.0}
\newcommand{\todo}[1]{\textcolor{red}{[[#1]]}}

\def\ignore#1{}

\newcommand{\cqcomment}[1]{\textcolor{blue}{[#1---cq]}}
\newcommand{\lycomment}[1]{\textcolor{red}{[#1---ly]}}

\newcommand{\jtcomment}[1]{\textcolor{orange}{[#1---jt]}}

\begin{document}

\title{Analyzing and Characterizing User Intent in Information-seeking Conversations}

\pagenumbering{gobble} 

\author{Chen Qu}
\affiliation{
	\institution{University of Massachusetts Amherst}
}
\email{chenqu@cs.umass.edu}

\author{Liu Yang}
\affiliation{
	\institution{University of Massachusetts Amherst}
}
\email{lyang@cs.umass.edu}

\author{W. Bruce Croft}
\affiliation{
	\institution{University of Massachusetts Amherst}
}
\email{croft@cs.umass.edu}

\author{Johanne R. Trippas}
\affiliation{
	\institution{RMIT University}
}
\email{johanne.trippas@rmit.edu.au}

\author{Yongfeng Zhang}
\affiliation{
	\institution{Rutgers University}
}
\email{yongfeng.zhang@rutgers.edu}

\author{Minghui Qiu}
\affiliation{
	\institution{Alibaba Group}
}
\email{minghui.qmh@alibaba-inc.com}

\begin{abstract}
    Understanding and characterizing how people interact in information-seeking conversations is crucial in developing conversational search systems. In this paper, we introduce a new dataset designed for this purpose and use it to analyze information-seeking conversations by user intent distribution, co-occurrence, and flow patterns. The MSDialog dataset is a labeled dialog dataset of question answering (QA) interactions between information seekers and providers from an online forum on Microsoft products. The dataset contains more than 2,000 multi-turn QA dialogs with 10,000 utterances that are annotated with user intent on the utterance level. Annotations were done using crowdsourcing. With MSDialog, we find some highly recurring patterns in user intent during an information-seeking process. They could be useful for designing conversational search systems. We will make our dataset freely available to encourage exploration of information-seeking conversation models.

\end{abstract}

%
%
%




\affiliation{%
  \institution{}
  \streetaddress{}
  \city{}
  \state{}
  \postcode{}
}

\maketitle

\section{Introduction}
\label{sec:intro}
Conversational assistants (CAs) such as Siri and Cortana are becoming increasingly popular. Users can issue simple queries and commands to a CA by voice to conduct single-turn QA or goal-oriented tasks, such as asking for weather and setting timers.
However, CAs are not yet capable of handling complicated information-seeking tasks which involve multiple turns of information exchange. These conversations are typically referred to as \textit{information-seeking conversations}, where the information provider (agent) provides answers to a query from an information seeker (user) and the agent modifies the answers based on user feedback.

To build functional and natural CAs that can reply to more complicated tasks we need to understand how users interact in these information-seeking environments. Thus, it is necessary to analyze and characterize user interactions and utterance intent. At CAIR\footnote{\url{https://sites.google.com/view/cair-ws/}} workshop at SIGIR'17, researchers indicated that there is a lack of conversational datasets to conduct studies. Therefore in this paper, we address this issue by collecting conversation data and creating the \textit{MSDialog}\footnote{The MSDialog dataset is available at~\url{https://ciir.cs.umass.edu/downloads/msdialog}} dataset. We present an analysis of user intent here, but MSDialog could also be used to conduct other dialog related tasks including response ranking and user intent prediction.

For effective analysis of user intent in an information-seeking process, the data should be multi-turn information-seeking dialogs. To support natural dialogs, conversational systems should be modeled closely to human behavior, thus the data should come from conversation interactions between real humans. As shown in Table~\ref{tab: datasets-compare}, we found that most existing dialog datasets are not appropriate for user intent analysis. The most similar data to ours is the Ubuntu Dialog Corpus (UDC), which also contains multi-turn QA conversations in the technical support domain. However, the user intent in this dataset is unlabeled. In addition, UDC dialogs are in IRC (Internet Relay Chat) style. This informal language style contains a significant amount of typos, internet language, and abbreviations. Another dataset, the DSTC 6 Conversation Modeling track data contains knowledge grounded dialogs from Twitter. However, this dataset contains scenarios where users do not request information explicitly, which do not fit the information-seeking narrative. Thus these datasets are not appropriate for user intent analysis. 
\def\semichecked{\checkmark\!\!\!\raisebox{0.4 em}{\tiny$\smallsetminus$}} 
\begin{table}[h]
\centering
\footnotesize
\caption{Comparison of related dialog datasets\ignore{ as for suitability of user intent analysis. The Twitter Corpus is private.}}
\vspace{-0.15in}
\label{tab: datasets-compare}
\begin{tabular}{@{}lllll@{}}
\toprule
Dataset                               & \begin{tabular}[c]{@{}l@{}}Multi-\\ turn\end{tabular} & \begin{tabular}[c]{@{}l@{}}Human-\\ human\end{tabular} & \begin{tabular}[c]{@{}l@{}}Information-\\ seeking\end{tabular} & \begin{tabular}[c]{@{}l@{}}User intent \\ label\end{tabular} \\ \midrule
DSTC 1-3~\cite{Henderson2015The}                             & $\checkmark$                                               &                                                        &                                                                &                                                              \\
DSTC 4-5~\cite{DSTC5}                               & $\checkmark$                                               & $\checkmark$                                                &                                                                &                                                              \\
Switchboard~\cite{Godfrey1997Switchboard}                          & $\checkmark$                                               & $\checkmark$                                                &                                                                &                                                              \\
{Twitter Corpus~\cite{Ritter:2010:UMT:1857999.1858019}} & {$\checkmark$}                        & {$\checkmark$}                         & {}                                        & {}                                      \\
DSTC 6 (2nd Track)~\cite{DSTC6_End-to-End_Conversation_Modeling}                 & $\checkmark$                                               & $\checkmark$                                                & $\checkmark\kern-1.1ex\raisebox{.7ex}{\rotatebox[origin=c]{125}{--}}$                                                       &                                                              \\
Ubuntu Dialog Corpus~\cite{Lowe2015The}                 & $\checkmark$                                               & $\checkmark$                                                & $\checkmark$                                                        &                                                              \\
\textbf{MSDialog}                              & $\checkmark$                                               & $\checkmark$                                                & $\checkmark$                                                        & $\checkmark$                                                      \\ \bottomrule
\end{tabular}
\end{table}

For open-domain chatting, it is common practice to train chatbots with social media data such as Twitter~\cite{Ritter:2011:DRG:2145432.2145500}. Similarly, real human-human multi-turn QA dialogs are the appropriate data for characterizing user intent in information-seeking conversations. In technical support online forums, a thread is typically initiated by a user-generated question and answered by experienced users (agents). The users may also exchange clarifications with the agents or give feedback based on answer quality. \ignore{Upon finishing the thread, the user (or the community) typically choose one or more answers that solve the original question as accepted answers to this thread. }Thus the flow of a technical support thread resembles the information-seeking process if we consider threads as dialogs and posts as turns/utterances in dialogs. We created MSDialog by crawling multi-turn QA threads from the Microsoft Community\footnote{\url{https://answers.microsoft.com}} and annotate them with fine-grained user intent types on an utterance level based on crowdsourcing on Amazon Mechanical Turk (MTurk)\footnote{\url{https://www.mturk.com/}}. \ignore{Although 
people behave differently when speaking and writing, the clear and formal language style of written dialogs in MSDialog should be a first step for analyzing user intent in information-seeking conversations. }

\begin{table*}[h]
\footnotesize
\centering
\caption{Descriptions and examples of user intent classes}
\vspace{-0.15in}
\label{tab:tagset}
\begin{tabular}{@{}llll@{}}
\toprule
Code & Label                 & Description                                                               & Example                                                                                \\ \midrule
OQ   & Original Question     & The first question by a user that initiates the QA dialog.                & If a computer is purchased with win 10 can it be downgraded to win 7? \\
RQ   & Repeat Question       & Posters other than the user repeat a previous question.                   & I am experiencing the same problem ...                                \\
CQ   & Clarifying Question   & Users or agents ask for clarification to get more details.                & Your advice is not detailed enough. I'm not sure what you mean by ... \\
FD   & Further Details       & Users or agents provide more details.                                     & Hi. Sorry for taking so long to reply. The information you need is ... \\
FQ   & Follow Up Question    & Users ask follow up questions about relevant issues.                    & Thanks. I really have one simple question -- if I ...                 \\
IR   & Information Request   & Agents ask for information of users.                                      & What is the make and model of the computer? Have you tried installing ... \\
PA   & Potential Answer      & A potential answer or solution provided by agents.                        & Hi. To change your PIN in Windows 10, you may follow the steps below: ... \\
PF   & Positive Feedback     & Users provide positive feedback for working solutions.                    & Hi. That was exactly the right fix. All set now. Tx!                   \\
NF   & Negative Feedback     & Users provide negative feedback for useless solutions.                    & Thank you for your help, but the steps below did not resolve the problem ...\\
GG   & Greetings/Gratitude   & Users or agents greet each others or express gratitude.                   & Thank you all for your responses to my question ...                    \\
JK   & Junk                  & There is no useful information in the post.                               & Emojis. Sigh .... Thread closed by moderator ...                       \\
O    & Others                & Posts that cannot be categorized using other classes.                     & N/A                                  \\ \bottomrule
\end{tabular}
\end{table*}

With this new dataset, we analyze the user intent distribution, co-occurrence patterns and flow patterns of large-scale QA dialogs. We gain insights on human intent dynamics during information-seeking conversations. One of the most interesting findings is the high co-occurrence of negative feedback and further details, which typically occurs after a potential answer is given. This co-occurrence pattern provides feedback about the retrieved answer and critical information about how to improve the previous answer. In addition, negative feedback often leads to another answer response, indicating that co-occurrence and flow patterns associated with negative feedback can be the key to iterative answer finding.

To sum up, our contributions can be summarized as follows. (1) We create a large-scale annotated dataset for multi-turn information-seeking conversations, which is the first of its kind to the best of our knowledge. We will make our dataset freely available to encourage relevant studies. (2) We perform in-depth data analysis and characterization of multi-turn human QA conversations. We analyze the user intent distribution, co-occurrence and flow patterns. Our characterizations also hold in similar data (UDC). Our findings could be useful for designing conversational search systems. 
\section{Related Work}
\label{sec:relatedwork}


Early conversational search systems through man-machine dialog include the THOMAS system by Oddy~\cite{Oddy1977Information}. It allowed users to conduct searches through dialogs. Belkin et al.~\cite{belkin1995merit} explored and demonstrated the justifiability of using information interaction dialogs to design the interaction mechanisms in IR systems. Shah and Pomerantz~\cite{Shah:2010:EPA:1835449.1835518} considered community QA as information-seeking processes and built models to predict answer quality. Radlinski and Craswell~\cite{Radlinski:2017:TFC:3020165.3020183} described a conceptual framework for conversational IR and the major research issues that must be addressed.

Recently, two observational studies captured how participants communicate and conduct searches in a voice-only setting~\cite{Trippas2017How, thomas2017MISC}. Both studies attempted to provide initial labeling for each utterance. \citet{Trippas2017How} analyzed the initial turns for patterns to classify with a qualitative analysis approach.\ignore{Their analysis showed the actions taken in a particular utterance. }
The MISC data~\cite{thomas2017MISC} came from similar experiments with data release including video, audio, and even emotions. Even though they offered valuable insights on how users conduct searches in a conversation, the data is not sufficient to perform a large-scale analysis and model training. \ignore{Abundant conversation data used in our analysis ensure that our characterization is more accurate and more consistent with actual human behavior.}

Also related to conversational search, Marchionini~\cite{Marchionini:2006:ESF:1121949.1121979} and ~\citet{DBLP:series/synthesis/2009White} addressed the importance of exploratory search, where the behavior of search is beyond a simple look up and more like learning and investigating. In this setting, the interpretation of user intent would rely heavily on the interactions between human and computer. This highlights the research need to characterize and understand user intent dynamics in information-seeking processes.

\section{The MSDialog Data}
\label{sec:data}
Our data collection contains two sets: the complete set and a labeled subset. Both will be publicly available. The complete set could be useful for unsupervised/semi-supervised model training.
The data used in the user intent analysis is the labeled subset.
In this section, we describe the three stages of generating MSDialog, which are data collection, taxonomy definition,  and user intent annotation.

\subsection{Data Collection}
\label{subsec:Dataset Collection}

We crawled over 35,000 dialogs from Microsoft Community, a forum that provides technical support for Microsoft products. This well-moderated forum contains user-generated questions with high-quality answers provided by Microsoft staff\ignore{, community moderators, article authors,} and other experienced users including Microsoft Most Valuable Professionals.

To ensure the quality and consistency of the dataset, we selected about 2,400 dialogs that meet the following criteria for annotation: (1) With 3 to 10 turns. (2) With 2 to 4 participants. (3) With at least one correct answer selected by the community. (4) Falls into one of the categories of Windows, Office, Bing, and Skype, which are the major categories of Microsoft products.

We observe that dialogs with a large number of turns or participants can contain too much noise, while dialogs with limited turns and participants are relatively clean. By choosing dialogs with at least one answer, we can use this dataset for other tasks such as answer retrieval. Also, by limiting the categories to several major ones, we can ensure language consistency across different dialogs, which is better for training neural models.


\subsection{Taxonomy for User Intent in Conversations}
\label{subsec:Classification Tagset}
We classify user intent in dialogs into 12 classes shown in Table~\ref{tab:tagset}.
Seven of the classes (\textit{OQ}, \textit{\textit{RQ}}, \textit{CQ}, \textit{FD}, \textit{PA}, \textit{PF}, \textit{NF}) were first introduced in FIRE'10\footnote{\url{https://www.isical.ac.in/~fire/2010/task-guideline.html}}. \citet{bhatia2012classifying} added the eighth class of \textit{Junk} as they observed a significant amount of posts with no useful information in their data (200 dialogs labeled with eight classes). 

We added four more classes to~\citet{bhatia2012classifying}'s taxonomy: \textit{Information Request}, \textit{Follow Up Question}, \textit{Greetings/Gratitude}, and \textit{Others}. We observed that agents' inquiries about user's version of software or model of computer is common in this technical support data and does not necessarily overlap with \textit{Clarifying Question}. \textit{Follow Up Question} is another utterance class in MSDialog as users sometimes expect agents to walk them step-by-step through the technical problem. \textit{Greetings/Gratitude} is quite common in the data. Finally, the \textit{Others} class is for utterances that cannot be classified with other classes. Note, each utterance can be assigned multiple labels because an utterance can cover multiple intent (e.g. \textit{GG+FQ}).

\subsection{User Intent Annotation with MTurk}
\label{subsec:Intention Annotation With Mechanical Turk}

\subsubsection{Procedure}
\label{subsubsec:procedure}
We employed crowdsourcing workers through MTurk to label user intent of each utterance using a set of 12 labels that is described in Section \ref{subsec:Classification Tagset}. The workers are required to have a HIT (Human Intelligence Task) approval rate of 97\% or higher, a minimum of 1,000 approved HITs, and be located in US, Canada, Australia or Great Britain. The workers are paid \$0.3/dialog.

In this annotation task, the workers are provided with a complete dialog. They are instructed to go through a table of labels with descriptions and examples before they proceed. For each utterance, the workers are tasked to choose all applicable labels that represent the user intent of the utterance and leave a comment if they choose the \textit{Others} label. 

\vspace{-0.08in}
\subsubsection{Quality Assurance}
\label{subsubsec:Quality Assurance}
To ensure the annotation quality, we employed two workers on each dialog. We calculated the inter-rater agreement using Fuzzy Kappa \cite{Kirilenko2016Inter} for this one-to-many classification task. We applied the threshold of 0.18 to filter the dialogs with too small Kappa scores, which reduced the number of dialogs by 9\%.





\section{Data Analysis \& Characterization}
\label{sec:analysis}

\subsection{Data Statistics}
\label{subsec:data-statistics}

The annotated dataset contains 2,199 multi-turn dialogs with 10,020 utterances. \ignore{It consists of technical support dialogs from four major categories of Microsoft products, which are Windows, Office, Bing, and Skype, and they are distributed roughly equally. }Table \ref{tab:properties} summarizes the properties of MSDialog. Each utterance has 1.83 labels on average.

\begin{table}[htp]
\footnotesize
\centering
\caption{Statistics of MSDialog}
\vspace{-0.15in}
\label{tab:properties}
\begin{tabular}{@{}lllll@{}}
\toprule
Items                      & Min & Max  & Mean   & Median \\ \midrule
\# Turns Per Dialog        & 3   & 10   & 4.56   & 4      \\
\# Participants Per Dialog & 2   & 4    & 2.79   & 3      \\
Dialog Length (Words)      & 27  & 1,467 & 296.90 & 241    \\
Utterance Length (Words)   & 1   & 939  & 65.16  & 47     \\ \bottomrule
\end{tabular}
\end{table}

\subsection{User Intent Distribution}

\begin{figure}[]
\centering
\begin{minipage}{0.22\textwidth}
  \centering
  \includegraphics[width=0.97\linewidth]{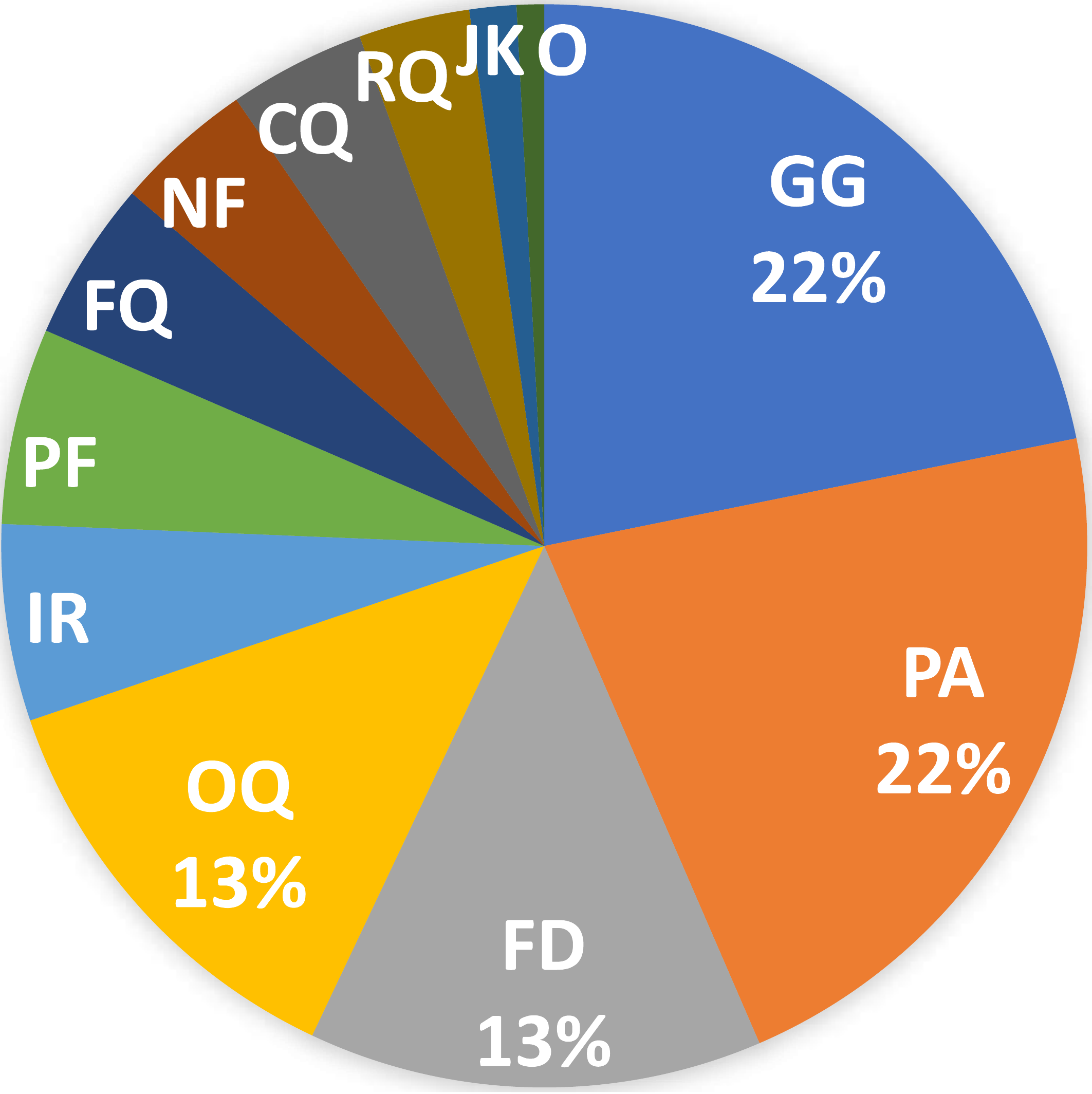}
  \vspace{-0.1in}
    \caption{Distribution}
	\label{fig:label-distribution}
\end{minipage}%
\begin{minipage}{0.28\textwidth}
  \centering
  \includegraphics[width=0.84\linewidth]{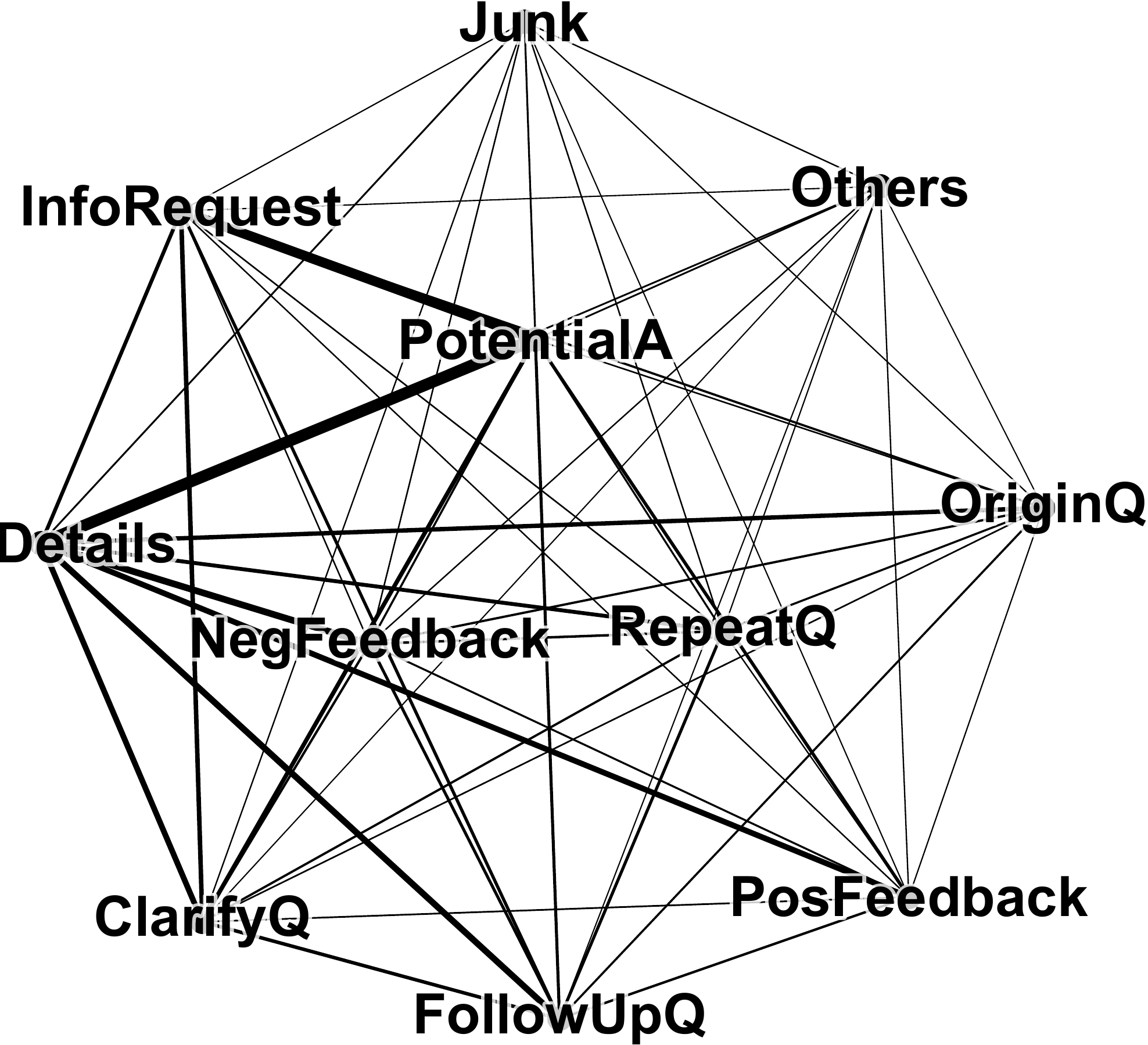}
  \vspace{-0.1in}
	\caption{Co-occurrence}
	\label{fig:label-co-occurruence}
\end{minipage}
\end{figure}

Figure~\ref{fig:label-distribution} shows the user intent distribution. Labels without a percentage are under 10\%. \textit{Greetings/Gratitude} and \textit{Potential Answer} are the most frequent labels\ignore{, contributing to almost half of the label occurrences}. This suggests that good manners and answers are at the center of human QA conversations. \textit{Repeat Question} is the most infrequent label except for \textit{Junk} and \textit{Others}, which is because the number of participants is limited to four. 

\subsection{User Intent Co-occurrence}
\label{subsec:label-co-occurrence}
Label co-occurrence in the same utterance can be useful for understanding user intent. Preliminary results indicate that the most frequent co-occurrence is between \textit{Greetings/Gratitude} and another label, suggesting good manners of forum users. 
Nevertheless, we removed \textit{GG} for the analysis later \ignore{due to the frequent occurrence. It allows us }to emphasize more on crucial user intent of information-seeking interactions.

The user intent co-occurrence graph with undirected edges weighted by co-occurrence count is presented in Figure \ref{fig:label-co-occurruence}. We observe that \textit{Potential Answer} often co-occurs with \textit{Further Details} or \textit{Information Request}. This indicates that agents tend to enrich possible solutions with details, or send \textit{Information Request}s in case the solutions do not work. Also, users tend to give \textit{Negative Feedback} with \textit{Further Details} to explain how the suggested answer is not working. In addition, \textit{Further Details} is observed to co-occur with \textit{Follow Up Question} or \textit{Clarifying Question}, suggesting that when people raise a relevant question, they tend to add details to them.

\subsection{User Intent Flow Pattern}
\label{subsec:markov}

\begin{figure}[h]
	\centering
	\includegraphics[width=0.47\textwidth]{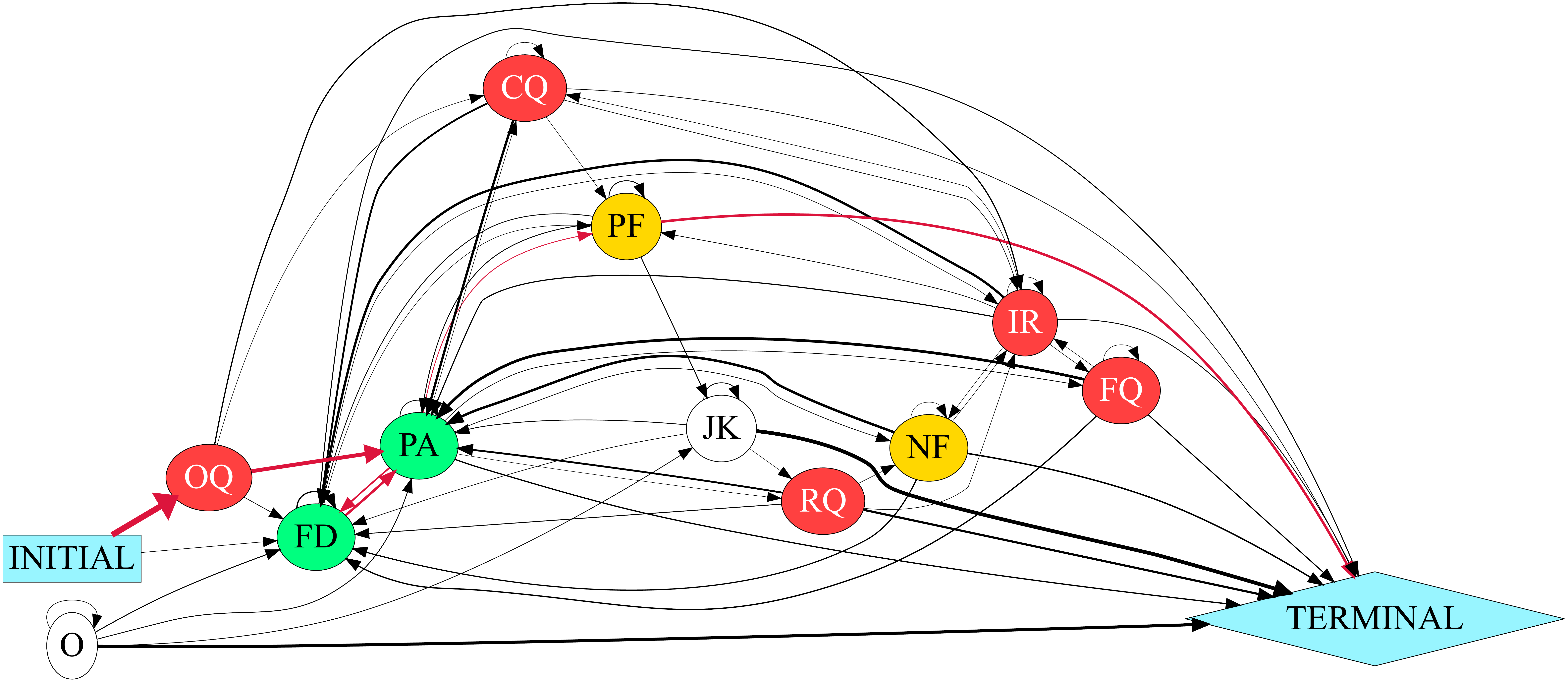}
	\vspace{-0.15in}
	\caption{Flow pattern with a Markov model. Node colors: red (questions), green (answer related), yellow (feedback). Edges are directed and weighted by transition probability.}
	\label{fig:markov}
\end{figure}

We use a Markov Model to analyze the flow patterns in the dialogs as shown in Figure \ref{fig:markov}. Because of the complexity and diversity of human conversations, many utterances are labeled with multiple user intent. We preprocess the traces (complete user intent flow in a dialog) with multiple labels by only using one label each time. For example, if we have a trace of "\textit{OQ}$\rightarrow$\textit{PA}+\textit{FD}$\rightarrow$\textit{PF}", we transfer it into two separate traces. The first one is "\textit{OQ}$\rightarrow$\textit{PA}$\rightarrow$\textit{PF}", and the second one is "\textit{OQ}$\rightarrow$\textit{FD}$\rightarrow$\textit{PF}". This preprocessing step can lead to a more concise model compared with using the original multi-labels as nodes. However, it does magnify some user intent nonproportionally. We alleviate the issue by only using dialogs that generate no more than 100 traces. This only filtered 30 dialogs.

In addition, we remove \textit{Greetings/Gratitude} because of the same reason described in Section \ref{subsec:label-co-occurrence}. Instead of simply hiding the \textit{GG} node from the final graph, we remove the occurrences of \textit{Greetings/Gratitude} if the utterance has multiple labels or change \textit{GG} to \textit{JK} if the utterance only has one label.

The flow pattern with a Markov model is presented in Figure~\ref{fig:markov}. As highlighted in the graph, a typical user intent transition path of MSDialog is "\textit{INITIAL}$\rightarrow$\textit{OQ}$\rightarrow$\textit{PA}$\rightarrow$\textit{FD}$\rightarrow$\textit{PA}$\rightarrow$\textit{PF}$\rightarrow$\textit{TERMINAL}". This represents the frequent user intent transition pattern in an information seeking process. We can make some observations from the graph : (1) In most cases, dialogs begin with an \textit{Original Question}, sometimes accompanied by \textit{Further Details}. (2) \textit{Original Question} tends to lead to \textit{Potential Answer} and \textit{Information Request}. (3) \textit{Information Request} and \textit{Clarifying Question} tend to lead to \textit{Further Details}. (4) \textit{Positive Feedback} tends to terminate the dialog while \textit{Negative Feedback} tends to lead to \textit{Potential Answer} or \textit{Further Details}. (5) Dialogs tend to end after \textit{Others} or \textit{Junk}. 


Besides the Markov transition graph, we use a different perspective to inspect the flow pattern by focusing on the user intent transition between turns in each dialog. We find that a quite significant flow path across turns is 
"\textit{INITIAL}$\rightarrow$\textit{OQ}$\rightarrow$(\textit{PA}$\rightarrow$\textit{FD}) $\times$3$\rightarrow$\textit{PA}$\rightarrow$\textit{PF}$\rightarrow$\textit{TERMINAL}". 
The "\textit{PA}$\leftrightarrow$\textit{FD}" circle pattern is typically caused by the "\textit{PA+IR}", "\textit{PA+CQ}", "\textit{NF+FD}" co-occurrences described in Section \ref{subsec:label-co-occurrence} and the "\textit{IR}$\rightarrow$\textit{FD}", "\textit{CQ}$\rightarrow$\textit{FD}", "\textit{NF}$\rightarrow$\textit{PA}" sequential relationship suggested in Figure \ref{fig:markov}.


\subsection{Comparison with Ubuntu Dialog Corpus}



Although UDC is less suitable for user intent analysis due to the informal language style, we investigate the characterizations of UDC and compare them to MSDialog since they are both in the technical support domain. We sampled 200 UDC dialogs and annotated user intent with MTurk using the same method with MSDialog. The informal language style of UDC may impact the annotation quality.
\ignore{Fortunately, 200 dialogs are sufficient to examine our findings. The comparison will be presented in terms of data statistics, label distribution, label co-occurrence, and flow pattern.}

\vspace{-0.08in}
\subsubsection{Statistics}
For this section, we present the statistics for UDC (complete set) and MSDialog (complete set) instead of the dialogs we sampled. As shown in Table~\ref{tab:compare-stats}, UDC dialogs have shorter utterances because of the informal language style. \ignore{\lycomment{More turns ? check the number ``7.71'' v.s. ``8.94'' in the table.}}

\begin{table}[ht]
\footnotesize
\centering
\caption{Statistics of UDC \& MSDialog (both complete sets)}
\vspace{-0.15in}
\label{tab:compare-stats}
\begin{tabular}{@{}lll@{}}
\toprule
Items                       & Ubuntu Dialog Corpus & MSDialog \\ \midrule
\# Dialogs                  & 930,000              & 35,000    \\
\# Utterances               & 7,100,000            & 300,000   \\
\# Words (in total)         & 100,000,000          & 24,000,000  \\
Avg. \# Participants        & 2                    & 3.18      \\
Avg. \# turns per dialog    & 7.71                 & 8.94     \\
Avg. \# words per utterance & 10.34                & 75.91    \\ \bottomrule
\end{tabular}
\end{table}

\vspace{-0.08in}
\subsubsection{Data Characterization}
\textit{Potential Answer} and \textit{Further Details} are the most significant user intent in UDC, which is consistent with MSDialog. Interestingly, the most common user intent in MSDialog, \textit{Greetings/Gratitude}, is quite rare in UDC. In addition, we observe the exact same top 5 label co-occurrences in UDC as described in Section \ref{subsec:label-co-occurrence}. Note that they are not necessarily in the same order. Finally, we found that the flow patterns observed in MSDialog also hold in UDC, except for the tendency from \textit{Positive Feedback} to \textit{TERMINAL}. This can be explained by the scarcity of \textit{Positive Feedback} in UDC. Although the UDC dialogs with informal language style are drastically different from the formal written style of MSDialog, the resemblance in user intent characterizations indicates that human QA conversations, regardless of the communication medium, follow similar patterns. 


\section{Discussion}
\label{sec:discussion}
In this section we discuss the limitation of our findings. The patterns we discovered are closely related to several design choices, including using dialogs from a well moderated forum in a specific domain. These choices were made to keep the setting as clean as possible as the research community is at an initial stage of this study. Although MSDialog does not cover every aspect of the highly diverse information-seeking conversations, it should be a first step to analyze and predict user intent in an information-seeking setting.

\section{Conclusions} 
\label{sec:conclusion}

In this paper, we create and annotate a large multi-turn question answering data for research in conversational search. We perform in-depth characterization and analysis of this data to gain insights on the distribution, co-occurrence and flow pattern of user intent in information-seeking conversations. We will make our dataset freely available to inspire future research. Future work will consider using neural architectures for user intent prediction tasks.



\ignore{\cqcomment{conference names to abbre}}

\begin{acks}
This work was supported in part by the Center for Intelligent Information Retrieval and in part by NSF grant \#IIS-1419693 and NSF grant \#IIS-1715095. Any opinions, findings and conclusions or recommendations expressed in this material are those of the authors and do not necessarily reflect those of the sponsor.

\end{acks}

\bibliographystyle{abbrv}
\bibliography{acmart} 

\end{document}